\documentclass[10pt,aps,prl,twocolumn,superscriptaddress]{revtex4-1}
\usepackage{graphicx}
\usepackage{color}
\usepackage{amsfonts}
\let\oldAA\AA
\renewcommand{\AA}{\text{\normalfont\oldAA}}

\begin{document}
\title{Photoemission Investigation of Oxygen Intercalated Epitaxial Graphene on Ru(0001)}
\author{S\o ren~Ulstrup}
\affiliation{Department of Physics and Astronomy, Interdisciplinary Nanoscience Center, Aarhus University,
8000 Aarhus C, Denmark}
\author{Paolo~Lacovig}
\affiliation{Elettra-Sincrotrone Trieste, S.S. 14 Km 163.5, 34149 Trieste, Italy}
\author{Fabrizio Orlando}
\affiliation{Physics Department, University of Trieste, Via A. Valerio 2, 34127 Trieste, Italy}
\affiliation{Paul Scherrer Institut, 5232 Villigen PSI, Switzerland}
\author{Daniel~Lizzit}
\affiliation{Elettra-Sincrotrone Trieste, S.S. 14 Km 163.5, 34149 Trieste, Italy}
\author{Luca~Bignardi}
\affiliation{Elettra-Sincrotrone Trieste, S.S. 14 Km 163.5, 34149 Trieste, Italy}
\author{Matteo~Dalmiglio}
\affiliation{Elettra-Sincrotrone Trieste, S.S. 14 Km 163.5, 34149 Trieste, Italy}
\author{Marco~Bianchi}
\affiliation{Department of Physics and Astronomy, Interdisciplinary Nanoscience Center, Aarhus University,
8000 Aarhus C, Denmark}
\author{Federico~Mazzola}
\affiliation{Department of Physics and Astronomy, Interdisciplinary Nanoscience Center, Aarhus University,
8000 Aarhus C, Denmark}
\author{Alessandro Baraldi}
\affiliation{Elettra-Sincrotrone Trieste, S.S. 14 Km 163.5, 34149 Trieste, Italy}
\affiliation{Physics Department, University of Trieste, Via A. Valerio 2, 34127 Trieste, Italy}
\affiliation{IOM-CNR, Laboratorio TASC, AREA Science Park, S. S. 14 Km 163.5, 34149, Trieste, Italy}
\author{Rosanna Larciprete}
\affiliation{CNR-Institute for Complex Systems, via dei Taurini 19, 00185 Roma, Italy}
\author{Philip~Hofmann}
\affiliation{Department of Physics and Astronomy, Interdisciplinary Nanoscience Center, Aarhus University,
8000 Aarhus C, Denmark}
\affiliation{philip@phys.au.dk}
\author{Silvano~Lizzit}
\affiliation{Elettra-Sincrotrone Trieste, S.S. 14 Km 163.5, 34149 Trieste, Italy}

\begin{abstract}
We study the formation of epitaxial graphene on Ru(0001) using fast x-ray photoelectron spectroscopy during the  growth process. The assignment of different C 1$s$ and Ru 3$d$ core level components and their evolution during the growth process gives a detailed insight into the graphene formation and the strongly varying graphene-Ru interaction strength within the large moir\'{e} unit cell. Subsequent intercalation of oxygen can be achieved at elevated temperature and the core level spectra show a conversion of the strongly corrugated to quasi free-standing  graphene, characterised by a single narrow C 1$s$ component. This conversion and the accompanying flattening of the graphene layer is also confirmed by x-ray photoelectron diffraction. The effect of oxygen intercalation on the electronic structure is studied using angle-resolved photoemission of the valence band states. For graphene/Ru(0001), the strong graphene-substrate hybridisation disrupts the $\pi$-band dispersion but oxygen intercalation fully restores the $\pi$-band with a strong $p$-doping that shifts the Dirac point 785~meV above the Fermi level. The doping of the system is highly tunable, as the additional exposure to rubidium can convert the carrier filling to $n$-type with the Dirac point 970~meV below the Fermi level.\\

\end{abstract}

\maketitle

\section{Introduction}
The epitaxial growth of graphene on metal surfaces has proven a viable route towards obtaining high-quality, large-area samples \cite{Sutter2008,Coraux:2008, Li:2009}, where detrimental effects on the electronic properties of graphene such as remote phonon scattering or charge puddles are avoided \cite{Chen:2008}. Some of the most studied epitaxial graphene-metal systems are graphene on Cu (GR/Cu) \cite{Li:2009,Walter:2011},  Pt(111) (GR/Pt) \cite{Preobrajenski2008GR},  Ir(111) (GR/Ir) \cite{Coraux:2008,Pletikosic:2009,Busse:2011},  Ru(0001) (GR/Ru) \cite{Marchini:2007,Sutter2008} and Ni(111) (GR/Ni) \cite{Varykhalov:2008}. One can order the graphene-substrate interaction in these cases from weak (GR/Cu, GR/Pt) to intermediate (GR/Ir) and strong (GR/Ru, GR/Ni). Tuning the strength of the interaction is essential for two properties: on the one hand, one wishes to have a sufficient interaction strength to avoid the growth of domains with rotational disorder, as observed for GR/Cu \cite{Walter:2011} and GR/Pt \cite{Preobrajenski2008GR}. On the other hand, a weak graphene-substrate interaction is highly desirable in order to preserve the electronic properties of graphene. In fact, in the strongly interacting cases the graphene $\pi$-states that make up the Dirac cone near the Fermi energy, $E_F$, can be hybridised with the electronic states of the substrate \cite{Busse:2011}, leading to the complete destruction of the Dirac cone \cite{Varykhalov:2008,Sutter2010}.

A possible solution of this dilemma is to grow graphene on a strongly coupling substrate and to subsequently decouple it by intercalation of atoms or molecules \cite{Varykhalov:2008}. Indeed, graphene on metal surfaces is highly prone to intercalation of a broad range of atoms and molecules that can not only decouple it from the substrate but also induce different degrees of doping  \cite{Varykhalov:2008,Sutter2010,Huang:2011,Larciprete2012,granas:2012,Petrovic:2013,Bignardi2017}. It has even been shown that sequential intercalation of different atoms can be used to grow insulators below the graphene to electrically decouple it from the metal surface \cite{lizzit:2012,Omiciuolo2014}.

In this paper, we focus on the GR/Ru system and the possibility to decouple the epitaxial graphene by oxygen intercalation. 
GR/Ru is one of the cases in which the graphene-substrate interaction is sufficiently strong to destroy the electronic structure of graphene near $E_F$. This system therefore requires intercalation \cite{Enderlein:2010,Sutter2010} or, similarly,  the growth of additional graphene layers \cite{Sutter:2009,Papagno:2012} in order to recover the expected electronic properties of graphene. 
GR/Ru and oxygen-intercalated graphene on Ru(0001) (GR/O/Ru) have been studied with x-ray photoelectron spectroscopy (XPS) as well as scanning probe and electron microscopy techniques \cite{Marchini:2007,Martoccia2008,Moritz2010,gyamfiinhomogeneous2011,Sutter:2013,Dong:2015,Voloshina:2016b}, revealing the strongly corrugated structure of the graphene and the relaxation of stress followed by wrinkle formation when decoupling the layer with oxygen \cite{Sutter:2013}.

Scanning tunnelling microscopy (STM) and spectroscopy (STS) studies have investigated the doping associated with the lifting of the graphene in GR/O/Ru, with somewhat contrasting results. One study found the Dirac point 480~meV below $E_F$ corresponding to a strong $n$-doping \cite{Jang:2013}. Another study interpreted the STS data with the help of density functional theory calculations of the graphene local density of states and concluded that a strong $p$-doping occurs with the Dirac point 552~meV above $E_F$ \cite{Voloshina:2016}. Angle-resolved photoemission spectroscopy (ARPES) experiments directly measure the $(E,k)$-dependent spectral function and have indeed also shown that oxygen intercalation can recover the Dirac cone of graphene in GR/O/Ru with the Dirac point $\approx$500~meV above $E_F$ \cite{Sutter2010}. There has also been a report of the graphene $\pi$-band's splitting into two states by ARPES, attributed to the varying interaction strength of graphene on Ru within the moir\'{e} unit cell. Remarkably, this splitting was not removed upon oxygen intercalation, even though the graphene corrugation should be greatly reduced \cite{Katsiev:2012}. 

Here we characterise the growth of GR/Ru and the oxygen intercalation, analyse the changes of the graphene-Ru interaction and investigate the coupling of the two materials within the moir\'{e} unit cell using XPS. We illustrate the change of graphene corrugation upon oxygen intercalation by x-ray photoelectron diffraction (XPD) and we present a careful characterisation of the valence band electronic structure of GR/Ru and oxygen-intercalated GR/Ru (with and without additional alkali metal doping) using high-resolution ARPES. 

\section{Experimental Methods}

The Ru(0001) sample was cleaned by repeated cycles of Ar$^+$ sputtering and annealings in O$_2$ atmosphere between 600 and 1000 K. The residual oxygen was removed by a final flash annealing up to 1500~K.

Graphene was grown by thermal decomposition of ethylene (C$_2$H$_4$) at 1100~K.
The precursor pressure was initially set to $5\times10^{-9}$~mbar \cite{Gunther2011} and successively increased in steps up to $5\times10^{-8}$~mbar to ensure a saturation coverage of the surface.
The resulting graphene layer showed the typical low energy electron diffraction (LEED) pattern with sharp moir\'{e} spots~\cite{Sutter2008}.

Oxygen intercalation was achieved by placing the sample in front of a custom-made O$_2$ doser and maintaining the O$_2$ background pressure at $5\times10^{-4}$ mbar.
The doser is a molybdenum tube of 6~mm diameter, placed at less than half a millimetre from the sample surface.
With this setup we estimate that the pressure at the sample surface is enhanced by about one order of magnitude compared to the background pressure.

The growth experiments and XPS measurements were carried out at the SuperESCA beamline of Elettra \cite{Baraldi:2003ab}.
High resolution XPS spectra of Ru 3$d$, C 1$s$ and O 1$s$ core levels were measured using a photon energy of 400 and 650~eV, respectively, with an overall energy resolution ranging from 40 to 100~meV. Unless stated otherwise, XPS data was collected at room temperature. The core level spectra were best fitted with Doniach-\u{S}unji\'{c} functions~\cite{Doniach1970} convoluted with a Gaussian distribution, and a linear background.
A Shirley background~\cite{Shirley1972} was subtracted from the Ru spectra before fitting.

XPD patterns were measured by collecting XPS spectra for more than 1000 different polar ($\theta$) and azimuthal ($\phi$) angles. 
For each of these spectra, a peak fit analysis was performed along the lines described above and the intensity $I(\theta$, $\phi)$ of each component resulting from the fit, \textit{i.e.} the area under the photoemission line, was extracted. The resulting XPD patterns are the stereographic projection of the modulation function $\chi$, which was obtained from the peak intensity for each of the polar and azimuthal emission angles $(\theta,\phi)$ as \[{\chi(\theta, \phi)} = \frac{I(\theta, \phi) - I_0(\theta)}{I_0(\theta)}\] where $I_0$($\theta$) is the average intensity for each azimuthal scan. A structural determination can be performed by comparing measured XPD patterns to multiple scattering simulations for a trial structure. Such patterns were simulated using the program package for Electron Diffraction in Atomic Clusters (EDAC) \cite{Garcia-de-Abajo:2001aa}.

The ARPES experiments were performed at the SGM-3 beamline of the synchrotron radiation source ASTRID \cite{Hoffmann:2004aa} with the sample temperature kept at 70~K.
The photon energy was 47~eV, with a total energy and $k$ resolution of 18~meV and 0.01~\AA$^{-1}$, respectively.  Electron doping of the oxygen intercalated graphene was achieved by evaporating rubidium from a SAES getter source.
The Dirac point was linearly extrapolated from the peak positions of momentum distribution curves (MDCs) corresponding to the left and right branches of the $\pi$-band in the energy range from 0.3 to 0.6~eV below the Fermi level for hole-doped graphene. For electron-doped graphene the $\pi^{\ast}$-band was extrapolated from 0.3 to 0.6 eV and the $\pi$-band from 1.6 to 1.9 eV, leading to two estimates of the Dirac point of which the average value is taken as the final value. 

\section{Results and Discussion}
\subsection{Core level analysis of graphene interaction with Ru(0001)}
\begin{figure} [t!]
\begin{center}
\includegraphics[width=0.45\textwidth]{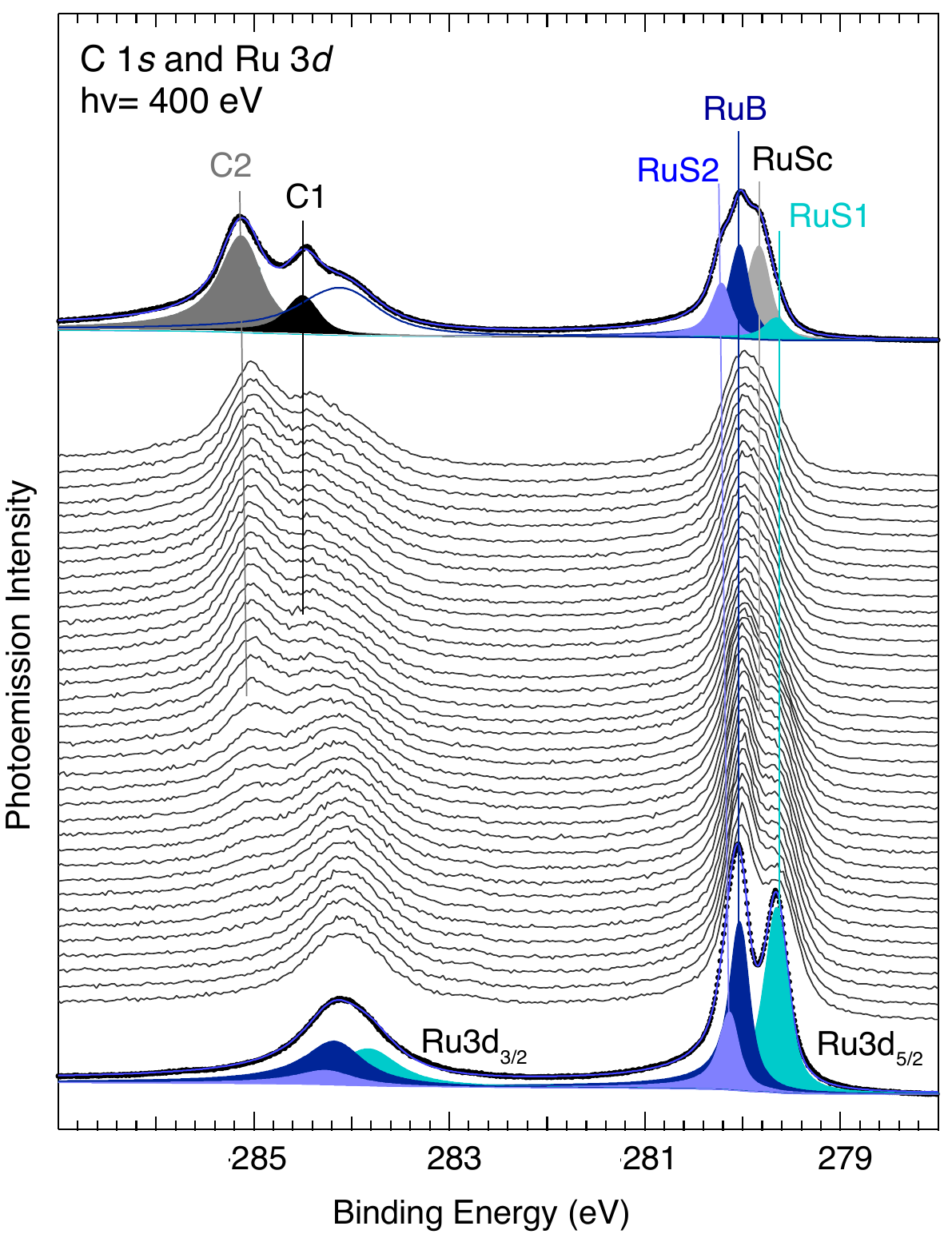}
\caption{Selection of fast-XPS C 1$s$ and Ru 3$d$ spectra measured during graphene growth at 1100 K. Bottom and top spectra were acquired at room temperature on the clean and graphene-covered Ru surface, respectively. For these spectra the deconvolution of the spectral components obtained from the fitting procedure is also shown (see text for details).}
\label{figure1}
\end{center}
\end{figure}

The XPS spectrum of the clean surface in the Ru 3$d$ core level region is shown in Fig.~\ref{figure1} (bottom).
The Ru 3$d$ spectrum shows the two spin-orbit split 3$d_{5/2}$ and 3$d_{3/2}$ components.
Three main contributions can be clearly distinguished in the Ru 3$d_{5/2}$ spectrum, namely RuB, RuS1 and RuS2, which belong to the bulk, first layer and second layer atoms, respectively. The RuB peak is found at 280.05~eV while RuS1 and RuS2 display a core level shift (CLS), that is the binding energy shift with respect to the bulk component, of -390~meV and 120~meV, respectively. These contributions are present also in the higher binding energy Ru 3$d_{3/2}$ peak, but they are rather broad due to a substantially shorter core-hole lifetime compared to that of the 3$d_{5/2}$ components~\cite{Lizzit2001}.
Also, the C 1$s$ core level that is observable after C$_2$H$_4$ exposure overlaps with the Ru 3$d_{3/2}$ core level region.
We therefore applied the following data analysis procedure in order to fit the Ru 3$d_{3/2}$ core level during graphene growth: The Ru 3$d_{5/2}$ peak was first fitted with the above described spectral contributions RuS1, RuS2 and RuB, plus the additional component RuSc stemming from the Ru surface atoms interacting with graphene. The Ru 3$d_{3/2}$ peak was then fitted with the same contributions but with a larger Lorentzian broadening, by applying the expected branching ratio of 1.5 between the areas of the spin-orbit split core levels Ru 3$d_{5/2}$ and 3$d_{3/2}$. 

Fig.~\ref{figure1} displays the evolution of the C~1$s$ state and the  Ru~3$d$ core levels during ethylene exposure at 1100~K. The high-energy resolution spectra measured at room temperature before and after graphene growth are shown in the bottom and top part of Fig.~\ref{figure1}, respectively. 

Two components appear in the C 1$s$ spectrum, C1 and C2 at 284.48 and 285.11~eV, respectively.
This double-peak structure is associated with a significant corrugation of graphene~\cite{Preobrajenski2008GR}, in accordance with previous investigations~\cite{Martoccia2008,Moritz2010}.
The C1 spectral component is related to the high regions (H) of the moir\'{e} unit cell~\cite{Preobrajenski2008GR}, while the high binding energy C2 peak stems from the strongly bonded parts of the graphene, which are closer to the substrate (L region)~\cite{Preobrajenski2008GR}.
It is important to note that the outlined picture is, however, only an approximate description of graphene chemisorption.
It has previously been pointed out that the peculiar shape of the C 1$s$ spectrum arises from a continuous distribution of non-equivalent C atomic configurations~\cite{Alfe2013}, leading to a broad distribution of binding energies within the moir\'{e} unit cell, rather than from two discrete regions of the graphene layer.
Similar results, achieved with the same combination of XPS characterisation and density functional theory calculations, were found for the graphene/Re(0001) interface~\cite{Miniussi2011}.

\begin{figure} 
\begin{center}
\includegraphics[width=0.49\textwidth]{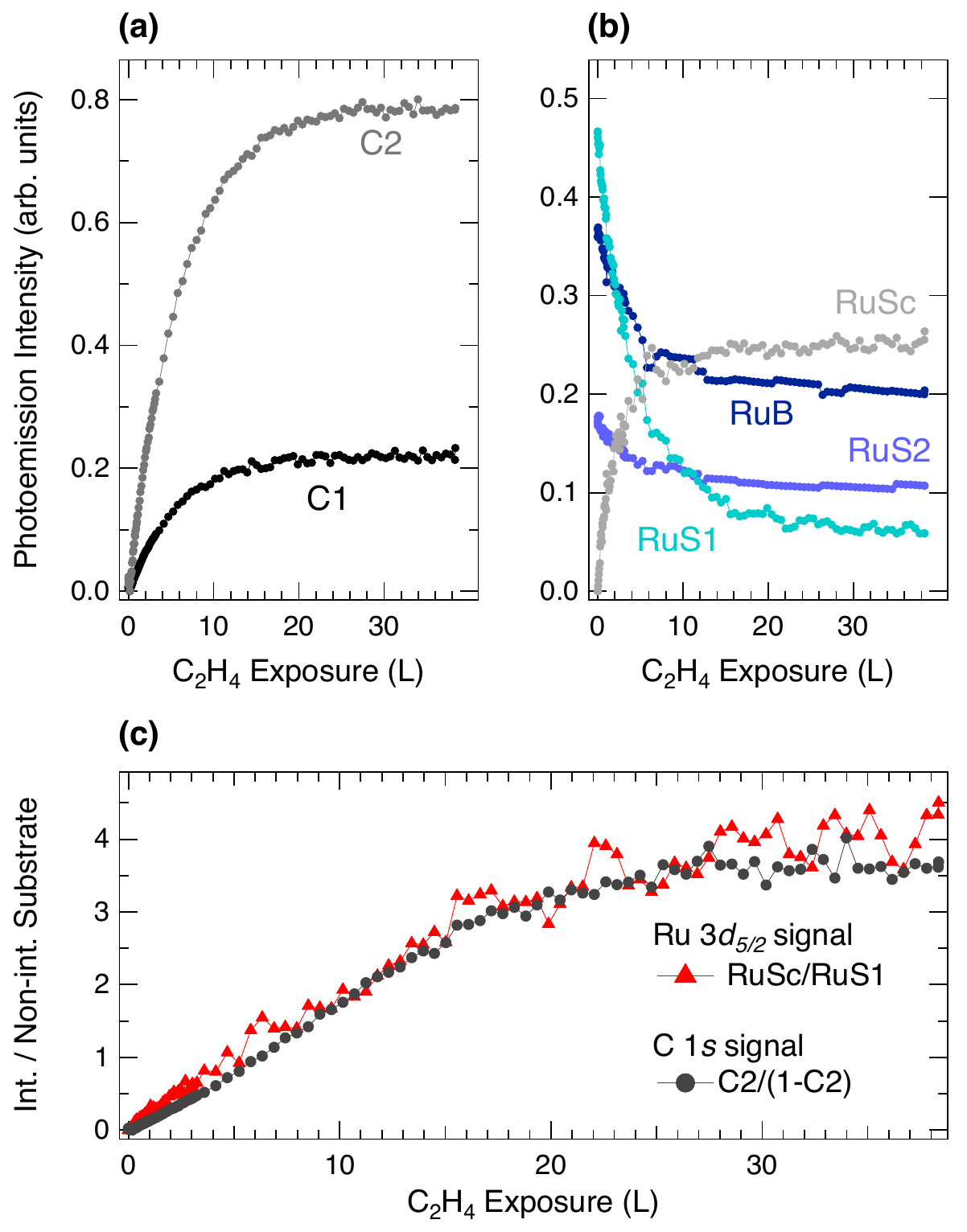}
\caption{Evolution of (a) C 1$s$, (b) Ru 3$d_{5/2}$ and (c) ratio between interacting and non-interacting Ru surface atoms as a function of ethylene exposure during graphene growth at 1100~K. (c) Data represented with red  triangles were obtained as the ratio between RuSc and RuS1 spectral components of the Ru 3$d_{5/2}$ core level, while dark-grey  circles have been calculated from C 1$s$ spectra as C2/(1-C2).}
\label{figure2}
\end{center}
\end{figure}

The evolution of the photoemission intensities of C1 and C2 as a function of the ethylene exposure is illustrated in Fig.~\ref{figure2}(a).
The two components grow with a constant ratio of $3.5\pm0.1$ throughout the whole process, eventually reaching saturation.
This finding is in agreement with the already proposed carpet-like growth, according to which the graphene islands are expected to grow in the down-direction of the staircase of the Ru substrate steps~\cite{Sutter2008,Loginova2009}.
Interestingly, STM measurements on graphene growth at low ethylene pressure, or high substrate temperature, highlighted a qualitatively different mechanism in which the graphene layer grows together with the underlying Ru terrace, in such a way that the graphene does not traverse the atomic step and keeps growing on the same Ru atomic plane~\cite{Gunther2011}. 
This mechanism, which implies the transport of ruthenium to increase the terrace size in the downhill direction, was observed also in the uphill direction, where it involves step etching~\cite{Gunther2011}.

Further details on the interaction between the graphene layer and the substrate can be inferred from the intensity changes of the  Ru 3$d_{5/2}$ spectral components shown in Fig.~\ref{figure2}(b).
As the ethylene exposure increases, the intensity of RuS1 is strongly reduced, whereas a new component, RuSc, grows with a CLS of -200~meV; it is straightforward to assign this spectral feature to Ru surface atoms interacting with the graphene layer. At the same time, the RuS2 component shifts by $\sim$70~meV towards higher binding energy.

Remarkably, the intensity of RuS1, even though decreased, does not vanish, as illustrated in Fig.~\ref{figure2}(b).
Because RuS1 is related to the clean Ru surface, this behaviour could indicate that, although saturation is reached, the graphene layer is not complete (saturation is evident from Fig.~\ref{figure2}(a)). However, this interpretation can be ruled out based on the system's behaviour upon subsequent exposure to oxygen. After exposing the surface to  $\sim$10$^5$~L of O$_2$ at room temperature, no O 1$s$ XPS signal could be detected, ruling out the existence of uncovered areas that would be immediately passivated by chemisorbed oxygen.

Most likely, the residual intensity of the RuS1 component corresponds to the surface atoms which are found below the elevated H part in the moir\'{e} unit cell, where the GR-Ru interaction is weak.
Therefore, these surface regions give rise to a core level shifted component very close to that of the clean surface, similarly to what has been previously shown for the GR/Ir interface~\cite{Larciprete2012}.
For the latter case, however, there are no H and L regions, but the whole graphene layer is weakly interacting with the substrate.
According to this interpretation, we developed the following picture: the H moir\'{e} regions present a C 1$s$ component at low binding energy and the corresponding Ru 3$d$ peak is quite unaffected by the presence of the graphene layer; in the L regions, however, the GR-Ru distance is smaller, the interaction is stronger and the C 1$s$ and Ru 3$d$ surface components are shifted towards higher binding energy.

Starting from this model, we have tried to correlate the intensity of the surface Ru 3$d$ with the C 1$s$ signals during the ethylene exposure.
According to the above assumption, the fraction of unperturbed Ru surface atoms consists of the sum of graphene-free areas and H regions, the L regions being the only portions of the substrate interacting with graphene.
The evolution of the ratio between interacting and non-interacting regions can be expressed using the intensity of either the C 1$s$ or Ru 3$d_{5/2}$ surface features which, in principle, should present the same behaviour. This is indeed found, as shown by the two curves in 
Fig.~\ref{figure2}(c).
The red  triangles represent the RuSc/RuS1 intensity ratio during ethylene exposure, representing the ratio of strongly interacting to weakly interacting plus clean surface areas. The corresponding curve for the carbon peak intensities is calculated as follows: C1 and C2 shall be the intensities of the two components as a function of ethylene exposure, normalized by the sum of C1 and C2 at saturation. Then C2 is the fraction of carbon atoms strongly interacting with the substrate and C1 is the fraction of weakly interacting carbon atoms. The fraction of clean surface is 1-(C1+C2). Therefore, 1-C2 corresponds to the sum of clean plus weakly interacting areas. The plot corresponding to RuSc/RuS1 is thus C2/(1-C2) which is shown as grey  circles.
The good agreement between the ratios calculated from these independent data sets (see Fig.~\ref{figure2}(c)) supports our interpretation.

\subsection{Evolution of GR-Ru coupling during oxygen intercalation}
In order to establish the best conditions for oxygen intercalation, the C~1$s$, Ru~3$d$ and~O 1$s$ core level spectra were measured for graphene exposed to molecular oxygen at different substrate temperatures.
Upon O$_2$ exposure at room temperature no O 1$s$ signal was detected; only the combination of elevated temperature and high local pressure allowed oxygen intercalation, confirming the completeness of the graphene layer also for GR on Ru(0001).    
The same behaviour was observed in the case of GR on Ir(111)~\cite{Larciprete2012}. 

\begin{figure} [b!]
\begin{center}
\includegraphics[width=0.49\textwidth]{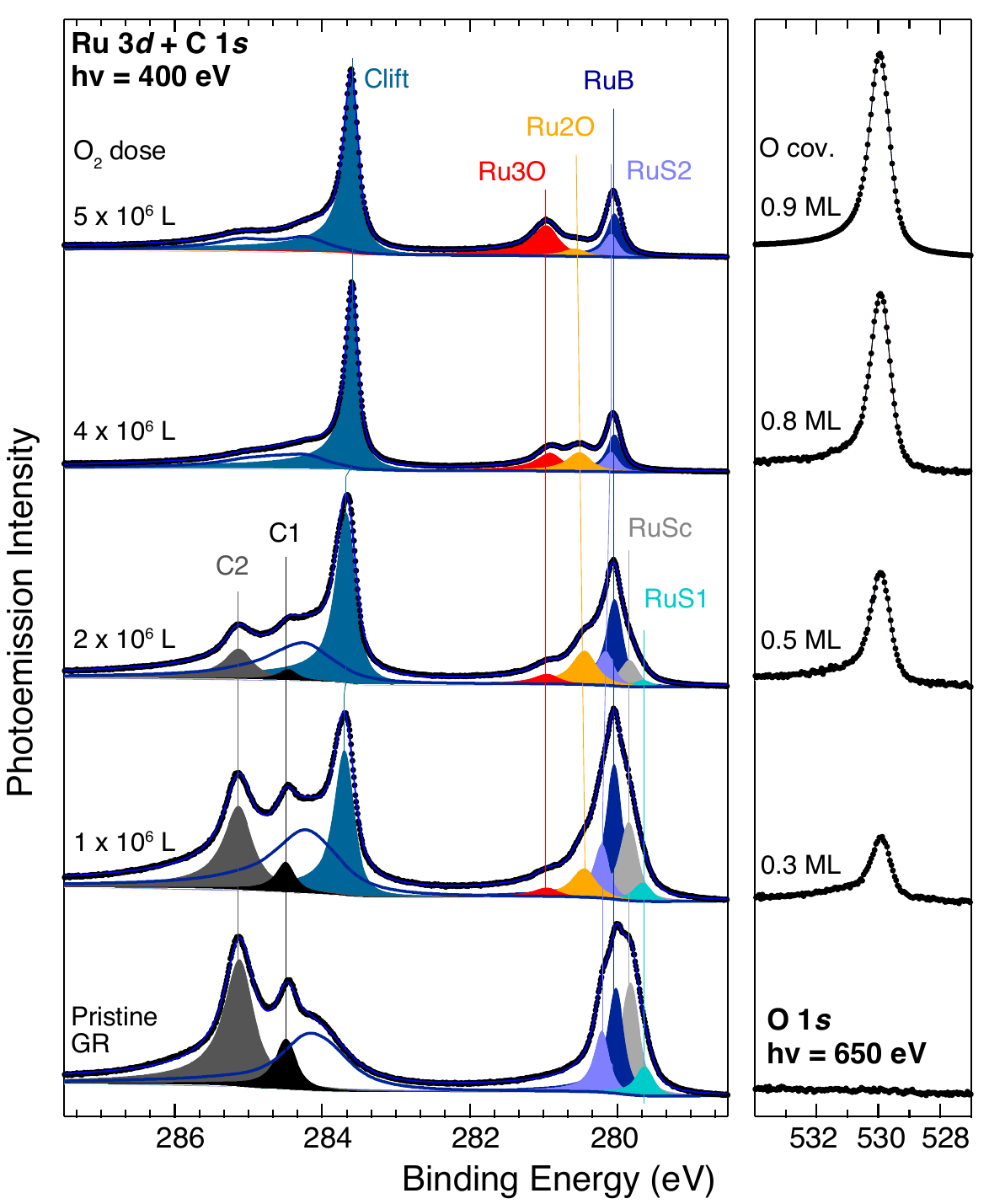}
\caption{Evolution of C 1$s$, Ru 3$d_{5/2}$ (left panel) and O 1$s$ (right panel) spectral components as a function of  oxygen exposure at a substrate temperature of 450~K. }
\label{figure3}
\end{center}
\end{figure}

\begin{figure} [t!]
\begin{center}
\includegraphics[width=0.5\textwidth]{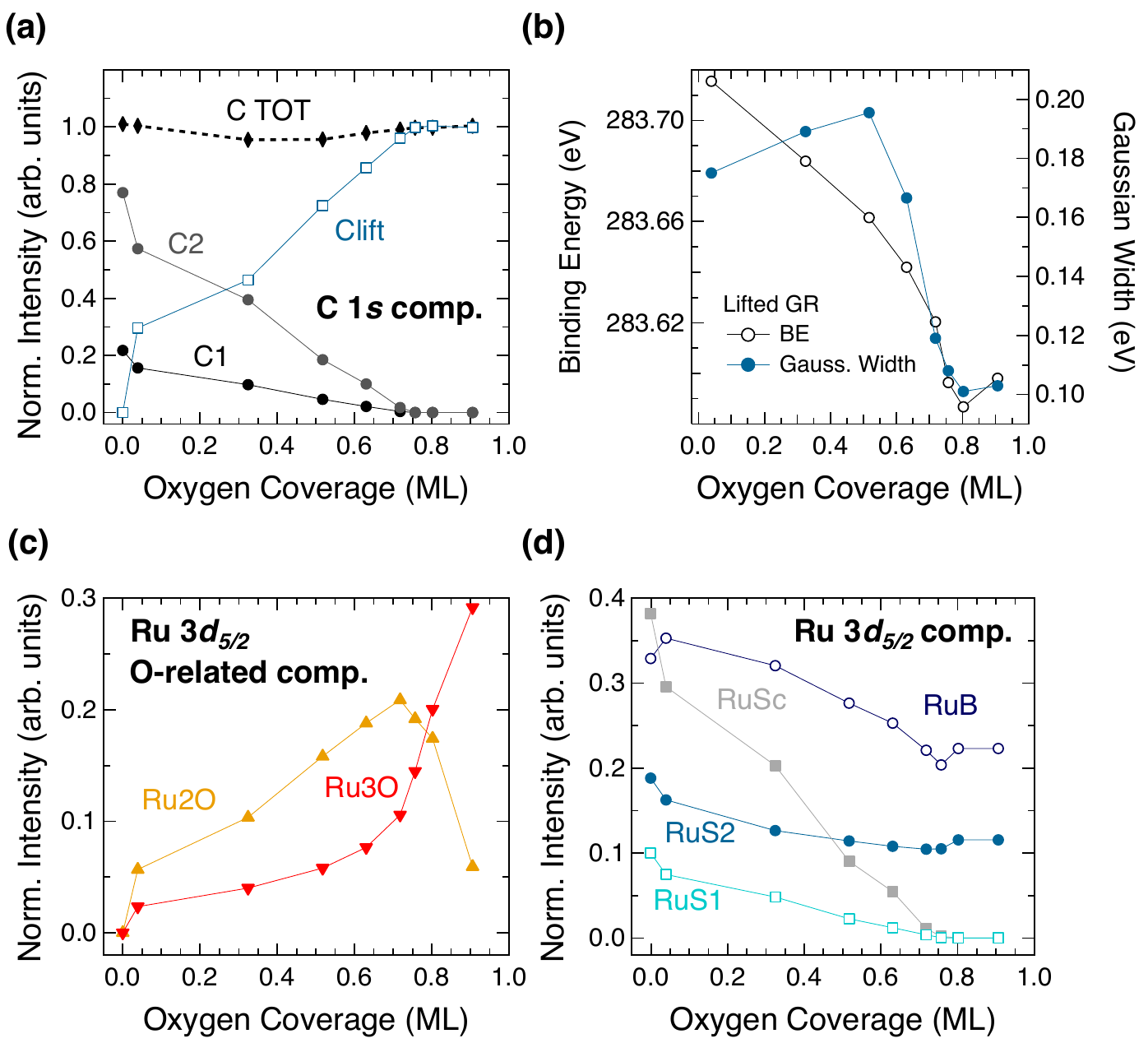}
\caption{Photoemission intensities of the (a) C 1$s$ and (c), (d) Ru 3$d_{5/2}$ components \textit{vs} oxygen coverage during the stepwise O intercalation process. (b) Binding energy shift and Gaussian width of the Clift component in the C 1$s$ core level \textit{vs} oxygen coverage.}
\label{figure4}
\end{center}
\end{figure}

At a substrate temperature of 420~K oxygen intercalates below graphene, consistent with previous results \cite{Dong:2015}, but for full intercalation and the consequent complete decoupling a temperature of 450~K is required. Going significantly above this temperature leads to oxygen deintercalation involving the removal of carbon.
XPS spectra measured during a stepwise O$_2$ exposure of graphene at 450~K are shown in Fig.~\ref{figure3}, while Fig.~\ref{figure4} shows the results of their analysis. The O coverage on the Ru substrate was calculated from the intensity of the O 1$s$ core level, assuming a final coverage of about 0.9~monolayers (ML), estimated by comparing the intensities of the oxygen-induced components in the Ru core level (Ru2O and Ru3O, see below) to those reported in Ref.~\cite{Lizzit2001}.

After an exposure of $\sim$10$^6$ L O$_2$, an intense O 1$s$ peak at 529.90 eV indicates the presence of 0.3~ML oxygen on the sample.
This is accompanied by strong modifications in the C 1$s$ spectrum: the C1 and C2 components lose intensity, while a new feature (Clift) grows, shifted by $\sim$750 meV towards lower binding energy with respect to C1.
As the oxygen exposure  increases, a further depletion of C1 and C2 is observed, accompanied by the intensity increase of the Clift component (Fig.~\ref{figure4}(a)), which undergoes a further 100 meV shift reaching the final binding energy of 283.60~eV and a line shape narrowing (the Gaussian width goes from 200~meV to $\sim$100~meV), suggesting an extended and uniformly lifted graphene layer (Fig.~\ref{figure4}(b)).

The effects of the oxygen exposure can also be observed in the line shape of the Ru 3$d_{5/2}$ core level.
The RuSc component progressively disappears, indicating a gradual reduction of the Ru surface areas interacting with graphene (Fig.~\ref{figure4}(d)).
In addition, two new spectral features appear at high binding energy, namely, Ru2O and Ru3O with a CLS of $\sim$410 and $\sim$900 meV, respectively (see Fig.~\ref{figure4}(c)).
These components were already measured in previous investigations of the O/Ru(0001) interface, and were assigned to Ru surface atoms bound to 2 and 3 oxygen atoms, respectively~\cite{Lizzit2001}.
The appearance of O-induced components in the Ru 3$d$ spectrum (see Fig.~\ref{figure4}(c)), together with the constant C 1$s$ intensity (see Fig.~\ref{figure4}(a)), suggests that an oxygen adlayer is forming underneath graphene, as already observed for graphene patches grown on Ru(0001)~\cite{Starodub2010,Zhang2009,Sutter2010}.

\subsection{Low energy electron diffraction and X-ray photoelectron diffraction}

\begin{figure} [t!]
\begin{center}
\includegraphics[width=0.45\textwidth]{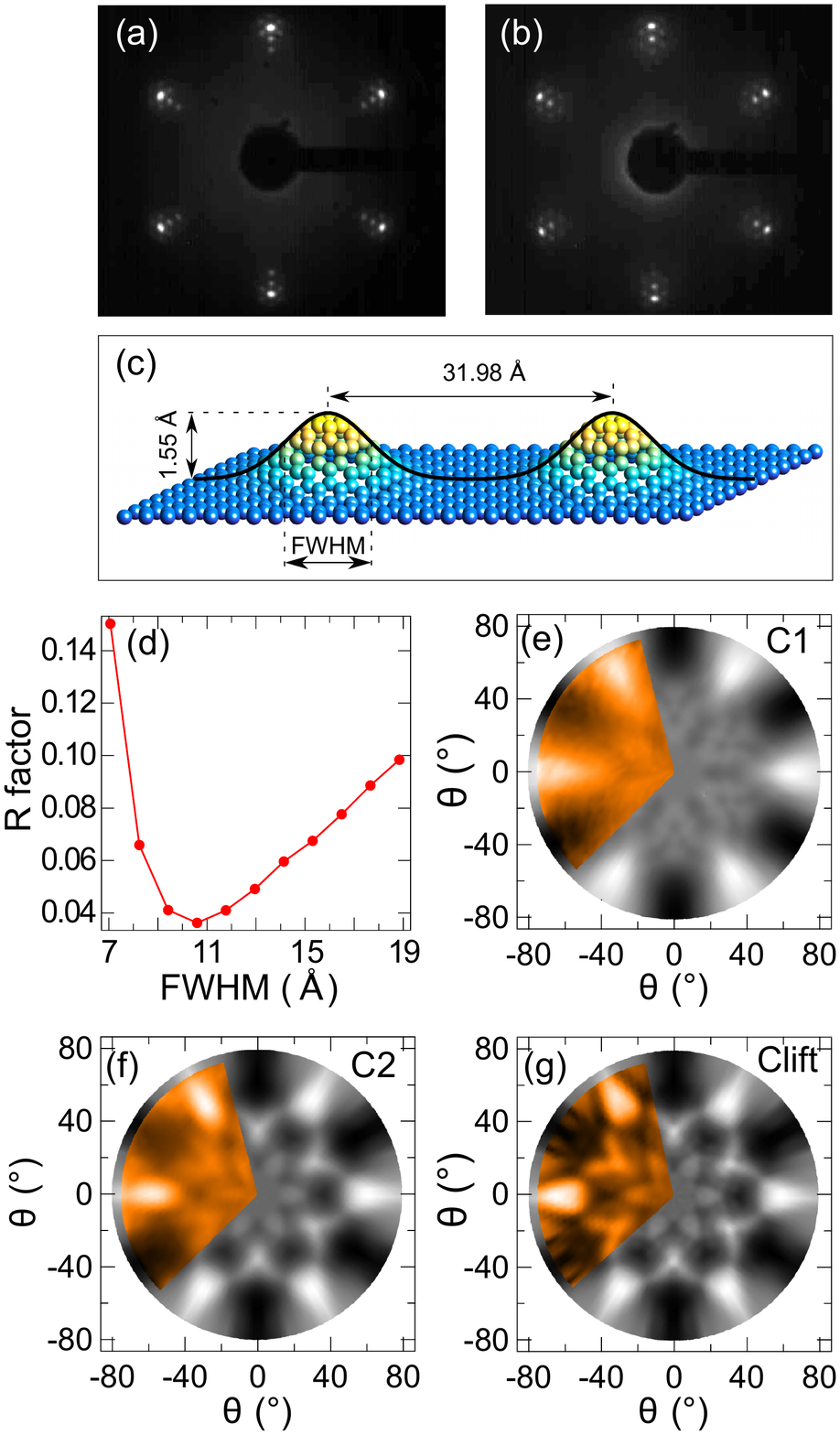}
\caption{(a), (b) LEED patterns from GR/Ru and GR/O/Ru, respectively. (c) Sketch of the structure used for the simulation of the XPD patterns of corrugated graphene. The distance between the two H regions corresponds to the ($13 \times 13$) moir\'{e} periodicity. (d) Reliability factor as a function of the Gaussian FWHM describing the corrugation, as explained in the text. (e),(f) Comparison between experimental data from the C1 and C2 components of the C $1s$ spectrum (coloured) and XPD simulations (greyscale) for the carbon atoms from the L and H regions, respectively, for the FWHM giving the minimum R-factor. (g) Corresponding XPD pattern of Clift for the oxygen-intercalated graphene and simulation for a flat graphene layer.}
\label{figureXPD}
\end{center}
\end{figure}

The structure of the graphene layer with and without intercalated oxygen has been characterised using LEED and XPD. The LEED patterns for the pristine and intercalated graphene layer on Ru(0001) are shown in Fig.  \ref{figureXPD}(a) and (b), respectively and they are particularly useful to asses the oxygen coverage in this system. Clearly, the intercalation process only leads to a minor change of the relative intensities of the Ru(0001), graphene and moir\'{e} LEED spots and not to the appearance of additional diffraction maxima. This is different from the situation described in Refs. \cite{Sutter:2013} and \cite{Voloshina:2016} which report the simultaneous formation of an oxygen $(2 \times 1)$ superstructure, characteristic for a 0.5~ML coverage. This difference is consistent with the significantly higher oxygen coverage close to 1~ML that is reached here.

The detailed structural changes in the graphene layer upon oxygen intercalation can be determined by XPD \cite{Woodruff:2007aa}. This technique is based on the emission angle-dependent modulations of the core level photoemission intensity from the different atoms in the layer. The intensity modulations arise from the length difference between individual scattering pathways from the emitting atom to the detector and the coherent interference of the scattered waves. The XPD modulations are thus directly reflecting the local structural environment of the emitting atom. 

To characterize the corrugation of graphene on Ru(0001), XPD data were collected from the C1 and C2 components of the C 1$s$ core level at a photon energy of 400~eV and compared to simulations for a free-standing graphene layer. Graphene buckling  was simulated  with an atomic cluster modelled by varying the out-of plane atomic coordinate according to a  2D periodic, isotropic Gaussian function with a full-width at half maximum (FWHM) ranging from 7 to 19~\AA, height of 1.55~\AA, and periodicity of ($13 \times 13$) graphene unit cells, as shown in Fig. \ref{figureXPD}(c)  \cite{Alfe2013}. XPD simulations for graphene related to L and H regions were performed by selecting as emitters all the atoms below and above half of the corrugation height, respectively. This model is based on the C $1s$ spectral distribution of the carbon atoms inside the moir\'{e} unit cell described in Ref.  \cite{Alfe2013}. 

The shape of the graphene corrugation was determined by minimising the Reliability factor \cite{Woodruff:2007aa} for the modulation function of the carbon atoms belonging to the H region as a function of the FWHM with a resulting excellent minimum R-factor of $\sim$0.04 for a FWHM of 10.6~\AA, as shown in Fig. \ref{figureXPD}(d). 
Fig. \ref{figureXPD}(e) and (f) show a comparison between the measured and calculated XPD patterns for the L and H regions, respectively, with the simulation performed using the FWHM value that minimises the R-factor.   Finally, Fig. \ref{figureXPD}(g) shows the comparison between the experimental XPD pattern of the C 1$s$ Clift component after oxygen intercalation at the final oxygen coverage of 0.9~ML and the simulation for a flat graphene layer. It can be noticed that the diffraction features are much better defined than those measured before oxygen intercalation, as expected for a situation with more carbon atoms in exactly equivalent geometric configuations. This finding together with the excellent agreement between experiment and simulation with an R-factor of 0.08 for oxygen intercalated graphene further supports the removal of the strong graphene corrugation upon oxygen intercalation \cite{Katsiev:2012,Voloshina:2016}.

\subsection{Valence band measurements}
\begin{figure} [t!]
\begin{center}
\includegraphics[width=0.49\textwidth]{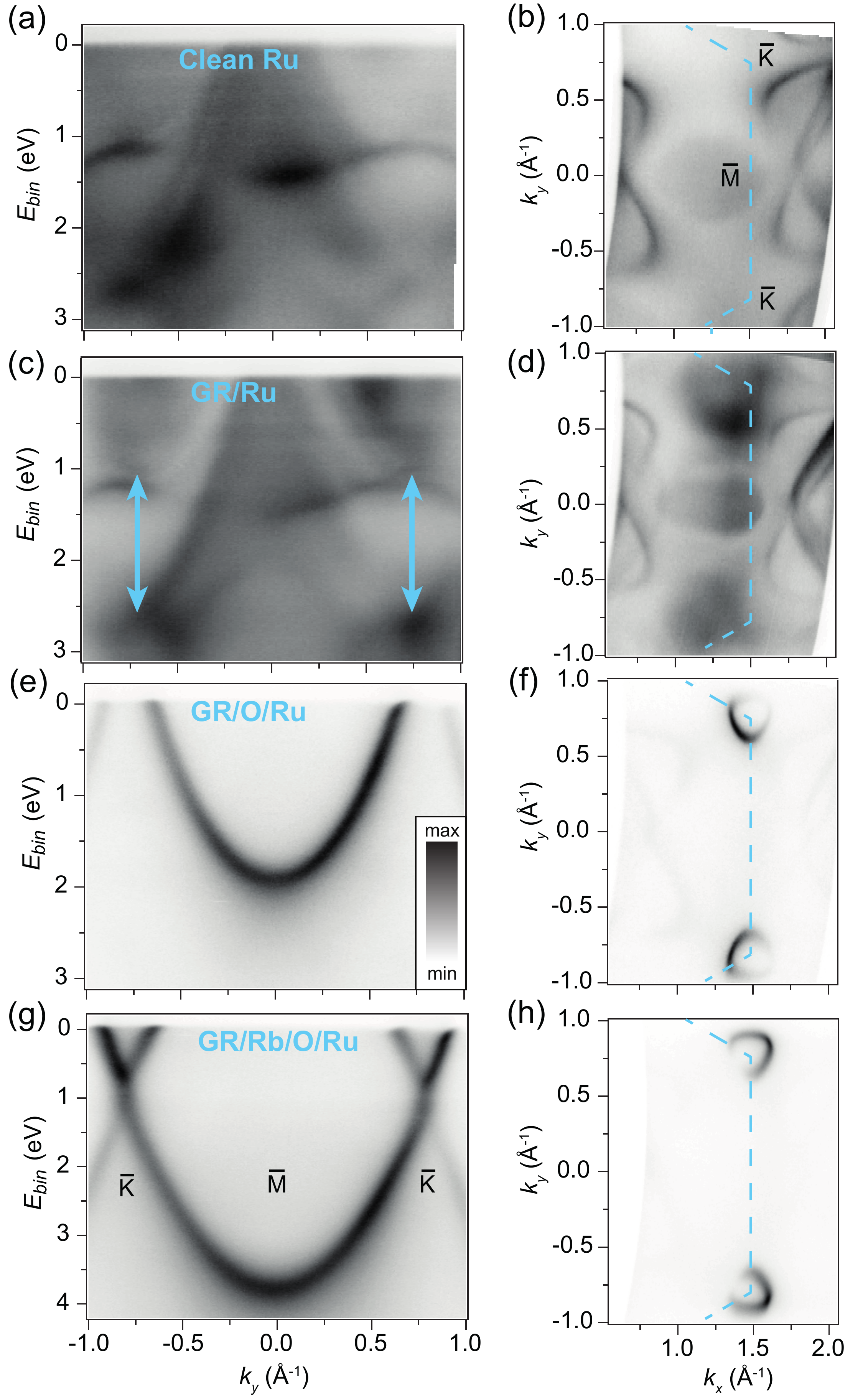}
\caption{Electronic structure measurements at a photon energy of 47~eV of (a)-(b) a clean Ru(0001) surface, (c)-(d) GR/Ru, (e)-(f) GR/O/Ru, and (g)-(h) GR/Rb/O/Ru. Panels on the left present cuts along the $\bar{\mathrm{K}}-\bar{\mathrm{M}}-\bar{\mathrm{K}}$ direction and panels on the right show constant energy cuts at the Fermi level with the graphene BZ marked by dashed lines. The arrows in (c) point at the onset of graphene related features.}
\label{figure5}
\end{center}
\end{figure}

ARPES measurements of the electronic structure of clean and graphene covered Ru are presented in Figs. \ref{figure5}(a)-(d). Panels~(a) and (c) show the dispersion along the $\bar{\mathrm{K}}-\bar{\mathrm{M}}-\bar{\mathrm{K}}$ direction of the graphene Brillouin zone (BZ) and panels~(b) and (d) present the measured segment of the Fermi surface with the graphene BZ outlined via dashed lines. Only subtle differences are visible between the two samples such as a broad feature around the $\bar{\mathrm{K}}$ points which disperses down to a binding energy of 1~eV in the case of the graphene-covered surface. The top of another set of broad bands can be distinguished from the background of Ru states around a binding energy of 2.5~eV. These broad features are interpreted as the $\pi^{\ast}$- and $\pi$-bands hybridized with Ru valence states and separated by a hybridisation-induced gap of around 1.5~eV (see double-headed arrows in panel~(c)). The broad and faint appearance of these bands is a consequence of the periodically varying chemical interaction with the substrate and the strong hybridisation between the Ru 4$d$ and the graphene $\pi$ states \cite{Brugger2009,Sutter2010}.

The oxygen intercalation procedure leads to the electronic structure shown in Figs. \ref{figure5}(e)-(f), where the photoemission intensity from bulk Ru states is strongly reduced while the characteristic hole-doped $\pi$-spectrum of oxygen intercalated graphene is clearly observed. We find that the Dirac point is located at a binding energy of ($-785 \pm 20$)~meV corresponding to $p$-doping with a carrier concentration of $4.5\times 10^{13}$ holes per cm$^2$. This observation confirms the oxygen intercalation of the full graphene layer on Ru, and that the Dirac cone can be recovered by switching off the hybridisation with the metal valence states \cite{Sutter2010}. Moreover, the higher oxygen coverage achieved here leads to an even stronger $p$-doping than reported previously \cite{Sutter2010,Voloshina:2016}. Note that no splitting of the Dirac cone is observed, in contrast to previous ARPES results for this system \cite{Katsiev:2012}. A possible cause of such splitting could be the presence of multiple graphene layers \cite{Papagno:2012}.

\begin{figure} [t!]
\begin{center}
\includegraphics[width=0.49\textwidth]{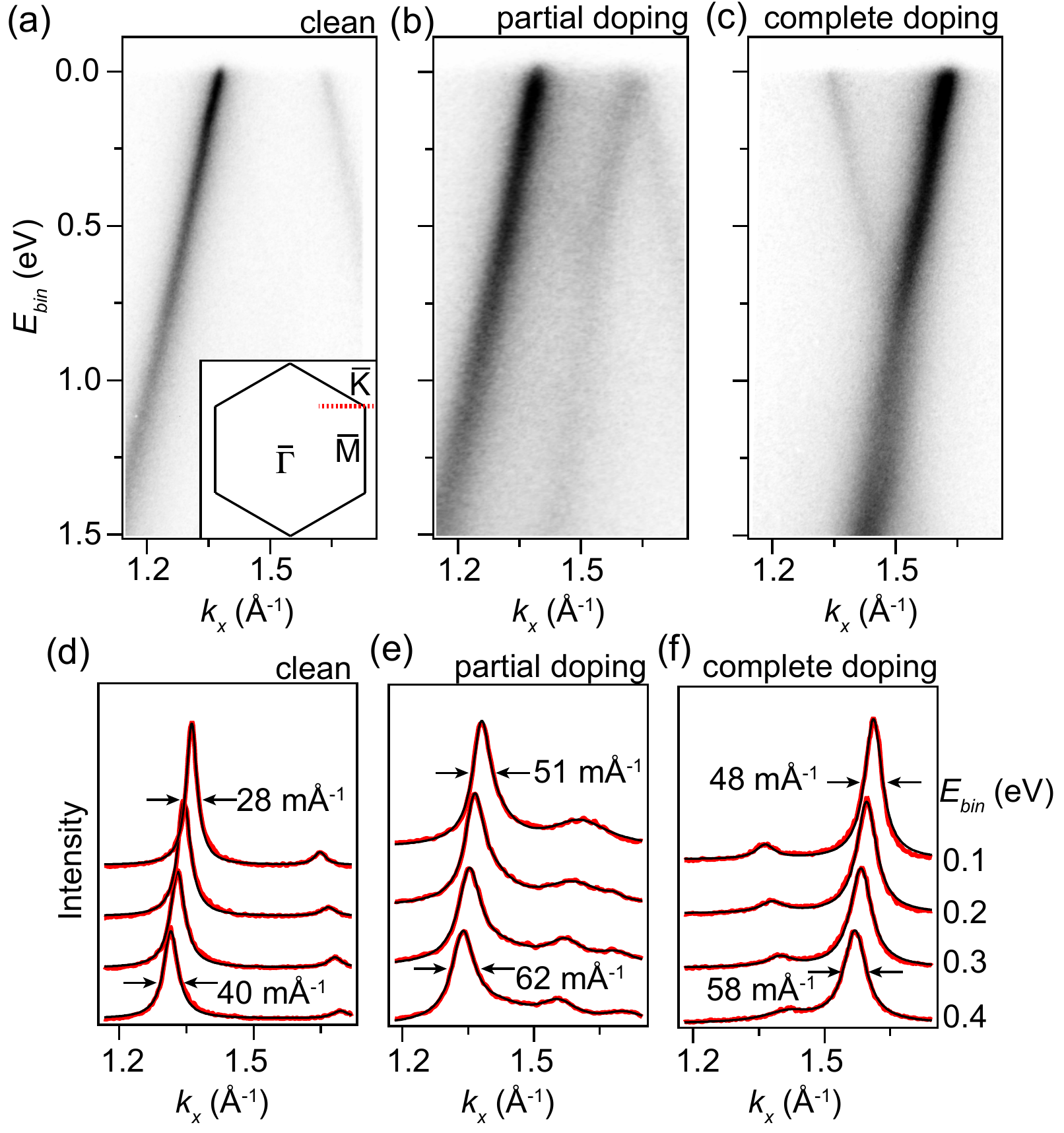}
\caption{Detailed ARPES measurements around the $\bar{\mathrm{K}}$ point of graphene for (a) clean GR/O/Ru, (b) partially Rb-doped GR/O/Ru and (c) fully Rb-doped GR/O/Ru (GR/Rb/O/Ru). The direction of the cut is shown via a dotted line on the BZ in the insert in (a). (d)-(f) Selection of MDCs (thick curves) and Lorentzian function fits (thin curves) in each case. The binding energies of the MDCs are given on the right in panel (f). The FWHM values of the most intense peak are provided at the highest and lowest binding energies of the fit and marked by arrows in each case.}
\label{figure6}
\end{center}
\end{figure}

We can further modify the electronic structure of the oxygen-intercalated system by doping it with alkali atoms. Figs. \ref{figure5}(g)-(h) show the electronic structure of GR/O/Ru when exposed to Rb atoms. The Dirac point is now located at a binding energy of ($970 \pm 100$)~meV corresponding to $n$-doping with a carrier concentration of $6.9\times 10^{13}$ electrons per cm$^2$. Note that the large error bar on the Dirac point position in this case results from the kink-like structure at the Dirac point, which may relate to electron-plasmon coupling, as observed in epitaxial graphene on silicon carbide \cite{Bostwick:2007}. This seamless switching from strong $p$- to strong $n$-doping of the graphene shows a high level of tunability of the Dirac cone in oxygen-intercalated graphene on Ru, similarly as demonstrated for graphene on Ir(111)~\cite{Ulstrup:2014e}.

Fig. \ref{figure6} shows more detailed measurements around $\bar{\mathrm{K}}$ for GR/O/Ru and for partial and full Rb doping. For each spectrum presented in Figs. \ref{figure6}(a)-(c) we analyze MDC cuts by fitting each branch with a Lorentzian peak and extract the full-width at half maximum (FWHM) for a range of binding energies, as shown in Figs. \ref{figure6}(d)-(f). In the case of graphene where the dispersion near $E_F$ is truly linear, the MDC FWHM is proportional to the electronic scattering rate arising from many-body interactions such as electron-electron, electron-phonon and electron-defect interactions \cite{Hofmann:2009ab}. We observe a FWHM value of 40~m$\AA^{-1}$ for GR/O/Ru at a binding energy of 0.4~eV, which is consistent with an intact high quality graphene layer where the electron-defect scattering rate is low \cite{Ulstrup:2014e}. Furthermore, the FWHM reduction to a value of 28~m$\AA^{-1}$ at a binding energy of 0.1~eV can be attributed to the varying electron-phonon coupling strength in this binding energy region, similarly as observed in quasi-free standing graphene on Ir(111) and on silicon carbide \cite{Johannsen:2013aa}. 

The ARPES intensity of partially Rb-doped GR/O/Ru reveals two sets of Dirac cones (see Fig. \ref{figure6}(b)), which are respectively $p$- and $n$-doped. The MDC fits in Fig. \ref{figure6}(e) include four Lorentzian peaks, which was required to obtain a good fit. A similar observation was made for Rb-doped oxygen intercalated GR/Ir and could be attributed to the coexistence of domains with adsorbed and intercalated Rb, resulting in different doping levels of graphene \cite{Ulstrup:2014e}. Assuming that a similar behavior leads to the two doped domains observed here on Ru(0001), the fully doped case in Fig. \ref{figure6}(c) would then correspond to complete Rb intercalation as only one strongly $n$-doped Dirac cone is observed. This is also clear from the MDCs in Fig. \ref{figure6}(f) where only two peaks are needed in order to obtain good fits. The higher FWHM values in panels~(e)-(f) compared to panel~(d) are likely caused by increased defect scattering due to the adsorbed or intercalated ionized Rb atoms.

\subsection{Discussion of intercalation mechanism}
The reported evolution of core levels and valence band depicts a system where the entire carbon network is detached from the substrate because of an oxygen layer intercalated between the graphene layer and the Ru surface. The situation is similar to the intercalation of a complete graphene layer on Ir(111)~\cite{Larciprete2012}.
The binding energy of the C 1$s$ component associated with the completely lifted graphene is the same (283.6 eV) on Ir and Ru.
However, there is a major difference in the behaviour of the C 1$s$ core level during oxygen intercalation for graphene grown on Ru(0001) and Ir(111).
For the Ir(111) substrate a rather continuous shift towards lower binding energy is observed (paralleled by an evident broadening in the intermediate stage) and ascribed to a charge transfer from carbon to the oxygen-covered Ir substrate, leading to a $p$-doped graphene.
In the case of Ru(0001), on the other hand, a new component is measured immediately after the first O$_2$ intercalation step, but the characteristic peaks of the pristine GR/Ru interface, although lowered in intensity, are still detected.
These findings can be rationalized by considering the different interaction of graphene with the two substrate.
Because of the weak GR-Ir interaction, the main effect of the oxygen intercalation is a net charge transfer which affects the layer as a whole.
In the Ru case, however, before the charge transfer can take place on the weakly interacting graphene, the strong GR-Ru chemisorption has to be weakened by the intercalated oxygen.

The mechanism driving the intercalation of oxygen below graphene can be assumed to be due to the penetration of O$_2$ molecules through pre-existing point defects and domain boundaries in the carbon lattice, as in the case of Ir.
Oxygen diffusion on the Ru surface can be reasonably expected to progress just like a water spill: the local O coverage is high close to the defects, so that the corresponding GR regions are detached and hole-doped. The O coverage decreases at larger distances and will eventually not be able to weaken the strong GR-Ru coupling, thus leaving pristine regions as witnessed by the presence of C1 and C2 components (Fig.~\ref{figure4}(a)).
Indeed, the Ru2O and Ru3O components start to develop already at very low coverages, wheras they are not present when oxygen is dosed on clean Ru(0001).
In the latter case, Ru2O and Ru3O develop from a coverage higher than 0.25 and 0.5 ML, respectively \cite{Lizzit2001}.
Subsequent oxygen exposures result in larger areas of decoupled graphene, paralleled by a more effective $p$-doping, as evidenced by the further binding energy shift experienced by the Clift component (Fig.~\ref{figure4}(b)).

In the case of the GR/Ir system, the progressive shift of the GR peak suggests a rather enhanced mobility of the oxygen atoms compared to that on GR/Ru, leading to a more uniform O coverage and, consequently, to a more efficient decoupling of the graphene layer even at a low oxygen concentration.
This is consistent with theoretical and experimental findings showing that graphene is more strongly bound to Ru(0001) than to Ir(111), thus making intercalation somewhat easier in the case of GR/Ir.
In this sense, the L graphene regions on Ru would act as `physical' barriers which hinder the migration of oxygen atoms, thus resulting in a lower oxygen mobility. 
This also correlates with the observed critical amount of oxygen of $\approx$ 0.75~ML  above which the lifting of GR on Ru is achieved (see Fig. \ref{figure4}(a)-(b)), which significantly exceeds the $\approx$ 0.45~ML oxygen coverage required on GR/Ir.

\section{Conclusion}
We have characterised the epitaxial growth and oxygen intercalation of graphene on Ru(0001) using \emph{in situ} XPS, tracking the C 1$s$ and Ru 3$d$ core level components associated with the H and L regions of the moir\'{e} unit cell. The strongly coupled graphene-Ru regions act as a barrier towards oxygen incorporation through the full carbon network. In spite of this, full intercalation could be achieved at a sample temperature of 450~K, leading to a C 1$s$ component (Clift) at a binding energy of 283.6~eV which is attributed to decoupled graphene. The $\pi$-band dispersion along the $\bar{\mathrm{K}}-\bar{\mathrm{M}}-\bar{\mathrm{K}}$ direction was measured by ARPES for the clean and intercalated systems. While the $\pi$ states are almost non-detectable for the non-intercalated system, narrow $\pi$ bands and strong $p$-doping are observed after oxygen intercalation. The doping can be widely tuned, even into the strong $n$-doping regime, by subsequent exposure to alkali atoms.

\section{Acknowledgements}
We gratefully acknowledge funding from VILLUM FONDEN through the Young Investigator Program (Grant. No. 15375) and the Centre of Excellence for Dirac Materials (Grant. No. 11744), as well as the Danish Council for Independent Research, Natural Sciences under the Sapere Aude program (Grant No. DFF-4002-00029) and the University of Trieste for the METAMATE project.

\end{document}